\documentclass{myws}

\begin{document}

\title{Hot Scalar Theory in Large \boldmath$N$: Bose-Einstein Condensation
\protect\footnote{{T\lowercase{alk presented at} SEWM 2000.}}}

\author{Peter Arnold}
\address{\rm reporting on work done in collaboration with}
\author{Boris Tom\'{a}\v{s}ik}
\address{
    Department of Physics,
    University of Virginia,
    P.O. Box 400714,
    Charlottesville, VA  22904-4714,
    USA
}

\maketitle

\def\Tc{T_{\rm c}}
\def\eff{{\rm eff}}
\def\grad{\mbox{\boldmath$\nabla$}}
\def\x{{\bf x}}
\def\y{{\bf y}}
\def\k{{\bf k}}

\abstracts{
I review
the Bose-Einstein condensation phase transition of dilute gases of cold atoms,
for particle theorists acquainted with methods of field theory
at finite temperature.
I then discuss how the dependence of the phase transition temperature
on the interaction strength can be computed in the large $N$ approximation.
}

\section{Phase Transitions in Hot Scalar Theories}

The standard example from particle physics of a scalar theory is the Higgs
sector of electroweak theory, which has a phase transition (or a crossover)
at a temperature of order the weak scale: say, a few hundred GeV or so.
Let's focus on pure scalar theory by imagining setting the gauge coupling
constant $g_{\rm w}$ to zero.  At finite temperature, the Higgs picks up a
thermal contribution $\#\lambda T^2$ to its effective mass, and the effective
potential becomes roughly of the form
$V(\phi) \sim m_\eff^2(T) \, \phi^2 + \lambda \phi^4$ with
$m_\eff^2(T) = - {\cal M}^2 + \#\lambda T^2$.
At sufficiently high temperature, the $\#\lambda T^2$ turns the potential
from being concave down at the origin to concave up, and so restores
the symmetry that is spontaneously broken at zero temperature.

Standard techniques for analyzing the phase transition between the hot,
symmetry-restored phase and the cold, symmetry-broken phase are as follows.
(i)
  Work in Euclidean time (for studying non-dynamical questions).
The Euclidean time direction becomes
periodic at finite temperature, with period $\beta=1/T$.
(ii)
  Near the second-order phase transition ($T \to \Tc$) of the purely scalar
model, the correlation length becomes infinite.  Large distance physics
becomes important, and there are large, non-perturbative, large-wavelength
fluctuations.
(iii)
  At large distances ($E_k \sim k \ll T$),
the compact Euclidean time direction decouples,
and one can reduce the original
Euclidean theory to a purely 3-dimensional effective theory of the
zero-Euclidean-frequency modes:
\begin {equation}
   S_{\rm eff} = \int d^3x\> \Bigl[
     |\grad\phi|^2 + m_\eff^2 |\phi|^2 + \lambda_\eff |\phi|^4
     + \cdots \Bigr] .
\end {equation}
(iv)
  Figure out what to do with the 3-dimensional theory (put it on a lattice,
or whatever).

\section {Today's Talk: Bose-Einstein Condensation}

The purpose of today's talk is to show that the exact same techniques
particle physicists use to study relativistically hot scalar theories can
also be used to study the Bose-Einstein condensation phase transition of
dilute gases of (for example) Rubidium atoms at $T \sim 0.1$ $\mu$K.
There's an identical three-dimensional effective theory to study
the non-perturbative long-distance physics ($E_k \ll T$) near the critical
temperature:
\begin {equation}
   S_{\rm eff} = \int d^3x\> \Bigl[
     |\grad\phi|^2 + r |\phi|^2 + u |\phi|^4
     + \cdots \Bigr] ,
\end {equation}
where I've switched to typical condensed-matter names ($r$ and $u$) for
the coefficients.  For a non-relativistic problem, $E_k \sim k^2/(2m)$,
and the long-distance
condition $E_k \ll T$ for the validity of this effective theory becomes
$k \ll \sqrt{2mT}$.

The fact that three-dimensional effective theories can describe second-order
phase transitions is old hat in condensed matter physics.  What's different
about the dilute atomic gas problem is that, just as in weakly coupled
relativistic problems, the {\it coefficients}\/ of that effective theory
are not phenomenological parameters but can be systematically matched to
the microscopic physics!  (This allows one to use the effective theory to
calculate more than just universal properties of the transition.)

\section{Non-Interacting Non-Relativistic Bose Gas}
The path integral corresponding to a
non-interacting, non-relativistic
Bose gas in an external potential $V(x)$ is
\begin {equation}
   Z = \int[{\cal D}\psi]\> e^{i\int dt\> L},
\end {equation}
\begin {equation}
   L = \int d^3x\>\psi^*\left(i\partial_t + {1\over2m}\, \nabla^2
           - V(x)\right) \psi
       + \mbox{(chemical potential term)}.
\end {equation}
where $\psi$ is a complex bosonic field and I've written the path integral,
for the moment, in real time rather than Euclidean time.
Why is this the path integral?
Note that the equation of motion, obtained by varying with respect to
$\psi^*$, is just the Schr\"odinger equation
$\left(i\partial_t+\nabla^2/2m-V(x)\right) \psi=0$.
The above path integral therefore describes the second quantization of
the Schr\"odinger equation: it describes arbitrary numbers of particles,
just like the standard path integral for QED describes arbitrary numbers
of photons.  In canonical quantization language, the field $\psi$
represents an operator
\begin {equation}
   \hat\psi(\x,t) = \sum_n \hat a_n \, \psi_n(\x) \, e^{-i\omega_n t},
\end {equation}
where the $\psi_n(\x)$ are eigenstates of the Schr\"odinger equation, the
$\omega_n$ are the corresponding eigen-energies, and the $\hat a_n$ are
the corresponding annihilation operators for particles in that mode.
If we specialize to the case where there is no external potential
$[V(\x)=0]$, then this becomes
\begin {equation}
   \hat\psi(\x,t) \to \int_\k \hat a_\k \, e^{i\k\cdot\x -i\omega_k t},
\end {equation}
which looks just like the quantization of field in terms of plane waves
that you're used to from relativistic quantum field theory.

Recall that in single-particle QM, $\psi^*\psi$ gives you the probability
density.
In second-quantized QM, the analogous statement is that
$\hat\psi^* \hat\psi$ gives you the number density, so
\begin {equation}
   \hat N = \int_\x \hat\psi^* \hat\psi
   = \sum_n \hat a_n^\dagger \hat a_n .
\end {equation}
To describe a system of particles with a given number density $n$, it's
convenient to use the grand-canonical ensemble and introduce a chemical
potential term $\mu N$ in the Hamiltonian or
Lagrangian.  So, our final Lagrangian for a free non-relativistic Bose gas
is
\begin {equation}
   L = \int d^3x\>\psi^*\left(i\partial_t + {1\over2m}\, \nabla^2
           + \mu
           - V(x)\right) \psi .
\end {equation}

\section {Interactions}

Now let's include a two-body potential $U(\x-\y)$ between the atoms
(which is adequate in the dilute limit).  Remembering that
$\psi^*\psi(\x)$ is the number density at $\x$, we need
\begin {equation}
   L_{\rm int} = -{1\over2}\int_{\x\y}
     \psi^*\psi(\x) \, U(\x-\y) \, \psi^*\psi(\y) .
\end {equation}
If $U$ is localized, then,
in the low-energy limit (wavelength $\gg a$), we can replace
it by an effective $\delta$-function:
\begin {equation}
   U(\x-\y) \to {4\pi a\over m} \, \delta(\x-\y),
\end {equation}
where $a$ is the {\it scattering length}\/ and can be measured for
a given type of atom.
For this substitution to be valid, the typical inter-particle spacing
$n^{-1/3}$ (where $n$ is the number density) should be large compared to
the scattering length.  In leading non-trivial order in this ``diluteness''
expansion $a \ll n^{-1/3}$, the Lagrangian for the interacting Bose gas
is then simply
\begin {equation}
   L = \int d^3x\>\left[\psi^*\left(i\partial_t + {1\over2m}\, \nabla^2
           + \mu
           - V(x)\right) \psi
        - {2\pi a\over m} (\psi^*\psi)^2 \right].
\end {equation}
I will henceforth focus on the case of zero external potential $V$.
In that case, this looks just like relativistic $\phi^4$ theory, but
with $\phi^*(-\partial_t^2)\phi$ replaced by $\psi^* i \partial_t \psi$.
In particular, $-\mu$ now plays the role that $m_\eff^2(T)$ did in the
introduction, and one can see that there will be a phase transition
depending on whether $-\mu(T) > 0$ (symmetry restored phase) or
$-\mu(T) < 0$ (symmetry broken phase).  In the latter case, the
expectation $\langle\psi\rangle \not= 0$ is what's known as the
Bose-Einstein condensate.%
\footnote{
   In graduate statistical mechanics, you learned for free particles
   that the condensate is the particles in the $\k=0$ mode.  The
   connection is that the contribution of $\k=0$ to
   the number density $n=\langle\psi^*\psi\rangle$ is
   $n_0 = \langle\psi\rangle^* \langle\psi\rangle$.
}

\section{Now Do the Usual}
To study the phase transition, go to Euclidean-time formalism.  As in
the introduction, note that all but the zero-Euclidean-frequency modes
$\psi_0$ decouple at large distance, leaving
\begin {equation}
   \int_0^\beta d\tau\> L_{\rm E} \to
   \beta 
   \int d^3x\>\left[\psi_0^*\left(-{1\over2m}\, \nabla^2
           - \mu_\eff
           + V(x)\right) \psi_0
        + {2\pi a\over m} (\psi_0^*\psi_0)^2 \right].
\end {equation}
[Corrections to this effective theory turn out to be higher order in the
diluteness expansion.]  This Lagrangian can be made to have standard
field-theory normalization for the kinetic term by rescaling
$\psi_0 = \phi \sqrt{2mT}$:
\begin {equation}
   \int_0^\beta d\tau\> L_{\rm E} \to
   \beta 
   \int d^3x\>\left[|\grad\phi|^2 +
           r |\psi|^2 + u |\phi|^4 \right] ,
\end {equation}
where $r=-2m\mu_\eff$ and $u=4\pi a m T$.
Note that the diluteness condition $a \ll n^{1/3}$ implies that the
coupling $u$ should be considered as ``small,'' since it is proportional
to $a$.  So, we have here an O(2) field theory in 3-dimensions
(since $\phi$ is complex) which is weakly-coupled at short distances.%
\footnote{
  More technically, this means weakly coupled at the short-distance scale
  $k \sim \sqrt{2mT}$ where it breaks down, where one wants to match it to
  the original theory.
}

\section{A Goal: Calculate \boldmath$\Tc$ as a Function of \boldmath$n$}

An interesting thing to try with this effective theory is to
calculate the correction, due to interactions, to the ideal gas result
for the Bose-Einstein condensation temperature $\Tc$.
Actually, it turns out to be technially slightly easier to calculate the
shift $\Delta n(T)$ in the critical density due to interactions, at fixed
temperature, rather
than the shift $\Delta T(n)$ in the critical temperature, at fixed density.
The two are easily related.  Then recall that the density
$n = \langle\psi^*\psi\rangle \sim \langle\phi^*\phi\rangle$.
If one were to do perturbation theory, the sort of diagrams one would
calculate for $\Delta n$ would then be of the form
\begin{center}
\epsfxsize=10pc
\epsfbox{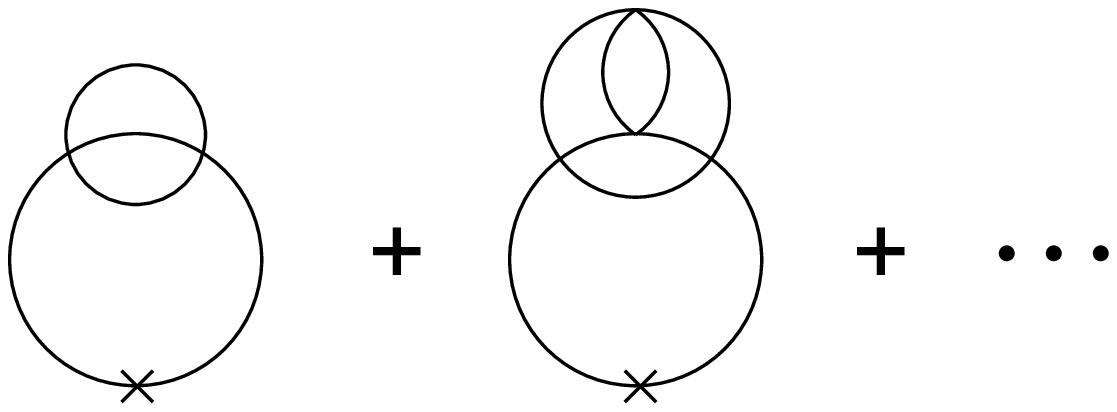}
\end{center}
where the cross represent the operator $\phi^*\phi$.  But perturbation
theory breaks down at the transition; so what to do?

One possible technique, implemented by Baym, Blaizot, and
Zinn-Justin,\cite{baym} is to try the large $N$ approximation for
solving the 3-dimensional theory, setting $N{=}2$ at the end.
At leading order in large $N$, the graphs which contribute to
$\langle\phi^* \phi\rangle \to \langle\vec\phi\cdot\vec\phi\rangle$
are
\begin{center}
\epsfxsize=3pc
\epsfbox{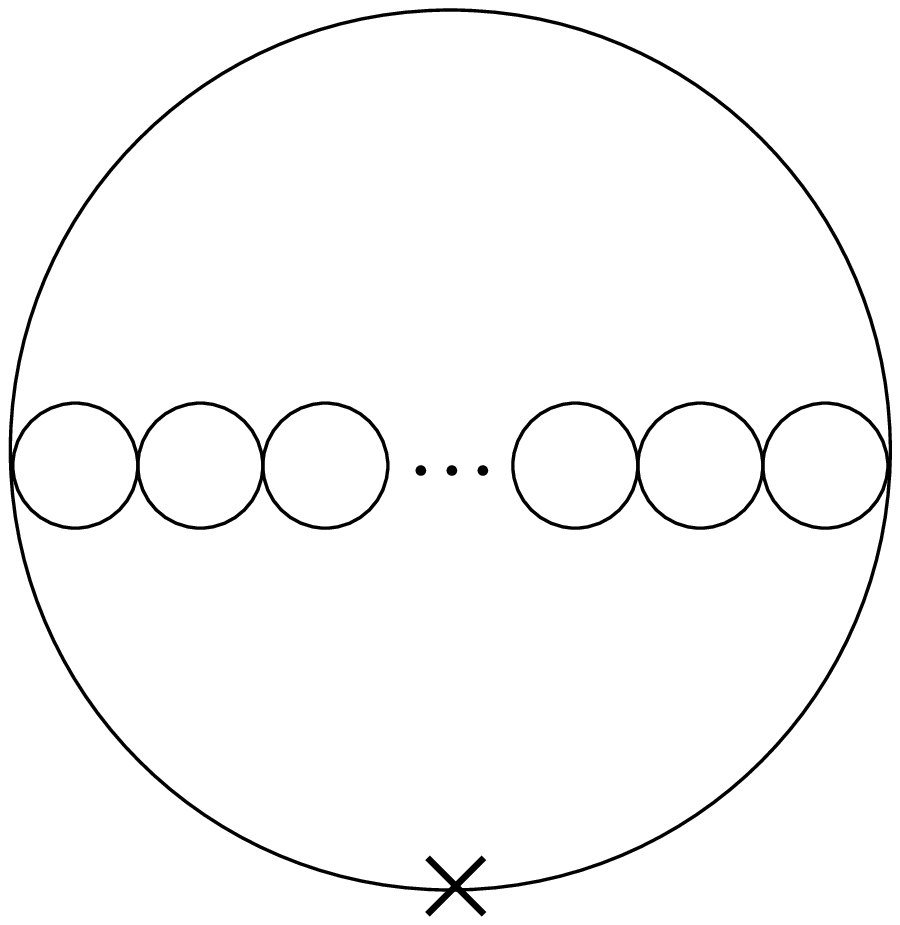}
\end{center}
Baym {\it et al.}\ find $\Tc = T_0 (1 + 2.33 \, a n^{1/3})$ plus higher
orders in $1/N$ and in $a n^{1/3}$.  Boris Tom\'{a}\v{s}ik
and I\cite{arnold}
have analyzed the next-order corrections in $1/N$ and find that they
change the coefficient 2.33 by only -26\% for $N{=}2$.
This correction is surprisingly small and suggests that large $N$
might not be too bad for $\Tc$!

\section*{Acknowledgments}
This work was supported by
the U.S. DOE, grant DE-FG02-97ER41027.

\end{document}